\documentclass{sty/acm_proc_article-sp}

\usepackage{url}
\usepackage{makeidx}  
\usepackage{amsfonts,epsfig,graphicx,enumitem,multicol} 
\usepackage{amsmath}
\usepackage{amssymb} 
\usepackage{mathrsfs}
\usepackage[lined,linesnumbered]
{algorithm2e}
\usepackage{verbatim} 
\usepackage{epic}
\usepackage{eepic}
\usepackage{latexsym}
\usepackage{graphics}
\usepackage{amsmath}
\usepackage{extarrows}
\usepackage{chemarr}

\newtheorem{example}{Example}

\begin{document}

\title{Efficient reduction of Kappa models by static inspection of the rule-set}

\numberofauthors{3} 

\author{
\alignauthor
Andreea Beica\\
       \affaddr{ENS Paris}\\
       \affaddr{Rue d'Ulm 45}\\
       \affaddr{Paris, France}\\
       \email{andrea.beica@ens.fr}
\alignauthor
Calin Guet\\
       \affaddr{IST Austria}\\
       \affaddr{Am Campus 1}\\
       \affaddr{Klosterneuburg, Austria}\\
       \email{calin.guet@ist.ac.at}
\alignauthor Tatjana Petrov\\
       \affaddr{IST Austria}\\
       \affaddr{Am Campus 1}\\
       \affaddr{Klosterneuburg, Austria}\\
       \email{tatjana.petrov@ist.ac.at}
}

\maketitle
\newcommand{\tanja}[1]{ \color{red} TP: #1 }

\def\MM{Enzymatic catalysis}
\def\mm{enzymatic catalysis}

\def\mmgen{generalized enzymatic catalysis}
\def\MMgen{Generalized enzymatic catalysis}

\def\fastdim{fast dimerization}
\def\Fastdim{Fast dimerization}

\def\specP{s}
\newcommand{\figref}[1]{Fig.~\ref{#1}}
\newcommand{\algref}[1]{Alg.~\ref{#1}}

\def\volume{N} 
\def\vel{v}
\def\Vel{V}

\newcommand{\secref}[1]{Section~\ref{#1}}
\newcommand{\tableref}[1]{Table~\ref{#1}}
\newcommand{\thmref}[1]{Theorem~\ref{#1}}

\def\N{{\mathbb{N}}}
\def\Rpos{{\mathbb R}_{\geq 0}}
\def\R{{\mathbb R}}
\def\indic{\mathsf{1}}
\def\pois{\xi}
\def\PP{\mathsf{P}}

\def\ra{\rightarrow}
\def\rA#1{\stackrel{\mathtt{#1}}{\ra}}

\def\noSubs{s}

\def\RS{reaction system}
\def\RBS{rule-based system}

\def\speciesSet{{\mathsf S}}
\def\reactionSet{\mathsf R} 

\def\spec{S_}
\def\reac{\mathsf{r}}
\def\specN{n}
\def\reacN{r}

\newcommand\cons[1]{{{a}}_{#1}} 
\def\zvec{\mathbf{z}}

\def\change{\nu}
\def\changev{\boldsymbol{\nu}}

\def\drate{k}
\def\srate{c}
\newcommand\consv[1]{{\mathbf{a}}_{#1}} 
\newcommand\prdnv[1]{\mathbf{a}'_{#1}} 
\newcommand\gainv[1]{\changev_{#1}}

\newcommand\gain[1]{\change_{#1}}
\newcommand\prdn[1]{{a}'_{#1}} 

\newcommand{\ka}[1]{\mathtt{#1}} 


\newcommand{\Rc}[1]{{R_{#1}}}
\newcommand\sprop[1]{\lambda_{#1}}
\newcommand\dprop[1]{\tilde{\lambda}_{#1}}

\def\xstov{\mathbf{\xsto}}

\def\noReac{m}
\def\noSpec{n}
\def\Nsp{n}
\def\Nreact{v}
\def\changevec{\vec{b}}

\def\stoich{\mathbf S}

\def\states{S}

\def\longrightharpoonup{\relbar\joinrel\rightharpoonup}
\def\longleftharpoondown{\leftharpoondown\joinrel\relbar}

\def\longrightleftharpoons{
  \mathop{
    \vcenter{
      \hbox{
    \ooalign{
      \raise1pt\hbox{$\longrightharpoonup\joinrel$}\crcr
      \lower1pt\hbox{$\longleftharpoondown\joinrel$}
    }
      }
    }
  }
}

\newcommand{\rates}[2]{\displaystyle
                  {\longrightleftharpoons^{#1}_{#2}}}

\def\procX{X}
\def\transient{p}
\def\pizero{\transient_0} 

\def\xdet{z} 
\def\xsto{x} 

\def\xdetv{\mathbf{z}}

\def\ystov{\mathbf{y}}

\def\specSpace{\N^{\specN}}
\def\patSpace{\N^{\patN}}
\def\patN{M}

\def\specN{n}
\def\reacN{r}

\newcommand{\conc}[1]{x_{#1}}
\def\xvec{\mathbf{x}}
\def\ratenew{\tilde{\rate}}
\newcommand{\class}[1]{[#1]}
\def\Xvec{X}

\def\der{\mathrm{d}}

\def\sp#1{\spec{#1}}
\def\xdet#1{[{#1}]}
\def\dx{\der x}
\def\dt{\der t}
\def\deriv{\frac{\der}{\dt}}
\def\consMatrix{P}
\def\changeMatrix{C}

\def\enz{E}
\def\subs{S}
\def\product{P}

\def\Xstov{\mathbf{X}}
\def\observs{{\cal O}}


\newcommand{\boundornot}{?} 
\newcommand{\bound}[1]{%
\ifnum\value{kappamode}=1
\_
\else
-\fi
}
\newcommand{\unirule}{ $\longrightarrow$ } 
\newcommand{\revrule}{ $\longleftrightarrow$ } 
\newcommand{\sitesep}{\adjust,}           
\newcommand{\agentsep}{~,~}    


\newcommand{\agenttextfont}[1]{\texttt{#1}}
\newcommand{\sitetextfont}[1]{\textit{#1}}

\newcommand{\statesymbol}{$\sim$}   
\newcommand{\bindingsymbol}{!}

\newcommand{\agentfigurefont}[1]{\texttt{#1}}
\newcommand{\sitefigurefont}[1]{\texttt{#1}}

\newcommand{\color}[2][]{}

\newcommand{\oderv}[2]{{\text{\productionprefix}}^{#1}_{#2}}
\newcommand{\odevar}[2]{{\text{\fragmentprefix}}_{#2}}
\newcommand{\odetime}{ \cdot }
\newcommand{\odeminus}{-}
\newcommand{\odeuniminus}{-}
\newcommand{\odeplus}{ + } 
\newcommand{\odeequal}{=}
\newcommand{\odeequalplus}{\stackrel{+}{=}}

\newcommand{\oderulename}[2]{\ruleprefix #2} 

\newcommand{\odebeforeequs}{}
\newcommand{\odeafterequs}{}
\newcommand{\odegrouptitle}[1]{\noindent Kinetic terms generated from rule #1:}
\newcommand{\odegroup}[2]{

\vspace*{5 mm}

\odegrouptitle{#1} 

\vspace*{5 mm}

#2}
\newcommand{\odesystem}{

\vspace*{5mm}

\noindent Differential equations: 

\vspace*{5mm}

}
\newcommand{\odeequ}[1]{\noindent $#1$ \smallskip\\}
\newcommand{\fragmentprefix}{$\mathcal{F}$}
\newcommand{\productionprefix}{$\mathcal{R}$}
\newcommand{\odediff}[2]{\overset{.}{\text{\fragmentprefix}_{#2}}}

\newcounter{dolabelrules}
\newcounter{rulenumber}
\newcommand{\putrulelabels}{\setcounter{dolabelrules}{1}}
\newcommand{\removerulelabels}{\setcounter{dolabelrules}{0}}
\newcommand{\ruleprefix}{} 
 
\newcommand{\changeruleprefix}[1]{\renewcommand{\ruleprefix}{#1}} 

\newcommand{\nextrule}{%
\ifnum \value{dolabelrules}=1 %
  \ruleprefix{%
\ifnum \value{rulenumber}<10 %
  0\therulenumber
  \else%
  \therulenumber
  \fi%
  {: }}%
  \fi%
\addtocounter{rulenumber}{1}%
}

\newcommand{\resetrulecounter}{\setcounter{rulenumber}{1}}
\newcommand{\nextrulebis}{\ruleprefix$_{\therulenumber}${:\ \ }\addtocounter{rulenumber}{1}}


\newcommand{\firstsep}{
\resetrulecounter%
\noindent%
\nextrule%
}
\newcommand{\rulesep}{  

\noindent%
\nextrule%
}             
\newcommand{\finalsep}{

}  

\newcounter{kappamode}
\newcommand{\figuremode}{\setcounter{kappamode}{1}}
\newcommand{\textmode}{\setcounter{kappamode}{0}}

\newcommand{\agentfont}[1]
{\ifnum \value{kappamode}=1
 \agentfigurefont{#1}%
 \else
 \agenttextfont{#1}%
  \fi} 
\newcommand{\sitefont}[1]
{\ifnum \value{kappamode}=1
  \sitefigurefont{#1}%
  \else
  \sitetextfont{#1}%
  \fi} 

\newcommand{\internalstate}[1]
{\def\arga{}
 \def\argb{#1}
 \textnormal{%
   \ifx\arga\argb
   {}
   \else%
   \textnormal{\statesymbol #1}
   \fi%
 }%
}%

\newcommand{\bondstate}[1]
{\def\arga{}
 \def\argb{#1}
 \textnormal{%
   \ifx\arga\argb
   {}
   \else%
   \textnormal{\bindingsymbol #1}
   \fi%
 }%
}%

\newcommand{\adjust}{}

\newcommand{\site}[1]{#1}


\newcommand{\agent}[2] 
{\def\arga{}%
\def\argb{#2}%
\ifx\arga\argb
\agentfont{#1}
\else%
{\agentfont{#1}
   \textnormal{$\left( \text{#2} \right)$}}%
 \fi}%

\newcommand{\btype}[2] 
{\ifnum \value{kappamode}=1
  {#2{}{}.#1{}}%
  \else
  {\text{#1{}$@$#2{}{}}}
  \fi} 
{}

\newcommand{\boundto}[3] 
{\ifnum \value{kappamode}=1
  {#2{}{}.#1{}}%
  \else
  {\text{#1$_{#2}${}$@$#3{}{}}}
  \fi} 
{}






\newcommand{\interface}{\Sigma}
\newcommand{\stateset}{\mathbb{I}}
\newcommand{\bindinglabel}{\mathbb{N}}
\newcommand{\strutfrac}[2]{\frac{\strut #1}{\strut #2}}
\newcommand{\rlhs}{lhs}
\newcommand{\rrhs}{rhs}
\newcommand{\refexpression}{E}
\newcommand{\var}{\mathcal{V}}
\newcommand{\sat}{\models}
\newcommand{\embed}[3]{#1 \lhd_{#3} #2} 
\newcommand{\dyn}{\mathbb{F}}
\newcommand{\auto}[1]{\embedset{#1}{#1}}
\newcommand{\nauto}[1]{|\auto{#1}|}
\newcommand{\logicand}{\wedge}
\newcommand{\logicor}{\vee}
\newcommand{\embedset}[2]{[#1,#2]}

 \newcommand{\Auf}{A}
\newcommand{\ABuu}{AB}
\newcommand{\ABpp}{{A^\star}\!B^\star}
\newcommand{\ABpu}{{A^\star}\!B} 
\newcommand{\ABup}{AB^\star}
\newcommand{\Apf}{{A^\star}}
 \newcommand{\Aub}{{AB^\diamond}}
 \newcommand{\Apb}{{A^\star}\!B^\diamond}

 \newcommand{\Buf}{B}
 \newcommand{\Bpf}{B^\star}
 \newcommand{\Bub}{{A^\diamond}\!B}
 \newcommand{\Bpb}{{A^\diamond}\!B^\star}

\newcommand{\xa}{n_{\Auf}}
\newcommand{\xas}{n_{\Apf}}
\newcommand{\xb}{n_{\Buf}}
\newcommand{\xbs}{n_{\Bpf}}
\newcommand{\xab}{n_{\ABuu}}
\newcommand{\xasb}{n_{\ABpu}}
\newcommand{\xabs}{n_{\ABup}}
\newcommand{\xasbs}{n_{\ABpp}}
\newcommand{\xx}[8]{(\xa#1,\xas#2,\xb#3,\xbs#4,\xab#5,\xasb#6,\xabs#7,\xasbs#8)}

\newcommand{\reactionbi}[5]{ & #2  & \xrightleftharpoons[#5]{#4} & #3 }

\newcommand{\rate}[1]{\mathtt{\tiny{k_{#1}}}}

\def\gr#1{\hat r} 
\def\emph#1{\textit{#1}}
\def\rar{\rightarrow}
\def\Rar{\Rightarrow}
\def\mbb#1{\mathbb{#1}}
\def\al{\alpha}
\def\ga{\gamma}
\def\eps{\epsilon}
\def\la{\lambda}
\def\sig{\sigma}
\def\Sig{\Sigma}
\def\lrar{\leftrightarrow}
\def\AR#1{\[\begin{array}{lllllll}#1\end{array}\]}
\def\EQ#1{\begin{eqnarray}#1\end{eqnarray}}
\def\doe{\downarrow_\emptyagent\hskip-.7ex}

\def\Ga{\boldsymbol\gamma}
\def\ga{\gamma}
\def\ep{\epsilon}

\def\I{\mathbf I}

\def\xAbs{\tilde{x}}
\def\lAbs{\tilde{l}}
\def\wAbs{\tilde{w}}

\def\WAbs{\tilde{\cal W}}

\def\PI{{\cal P}}

\def\L{{\mathcal L}}
\def\S{{\mathcal X}}
\def\M{{\mathcal M}}

\def\x{{\mathbf x}}
\def\Tauu{\tau_{\IReal}}
\def\tildtauu{\tilde{\tau}_{\IReal}}
\def\tildetset{\mathcal{T}_{\IReal}(\tilde{\mathcal{W}})}
\def\tset{\mathcal{T}_{\IReal}(\mathcal{W})}
\def\Ac{{\mathcal A}}

\def\x{{\mathbf x}}
\def\a{{\mathbf a}}
\def\b{{\mathbf b}}
\def\F{{\mathbf F}}
\def\G{{\mathbf G}}
\def\k{{\mathbf k}}
\def\s{{\mathbf s}}
\def\f{{\mathbf f}}

\def\Q{{\mathbf Q}}
\def\A{{\mathscr A}}

\def\myalpha{\bm{\alpha}}
\def\mybeta{\bm{\beta}}

\def\c{{\mathbf c}}

\def\R{{\mathbb R}}

\def\ra{\rightarrow}
\def\rA#1{\stackrel{{#1}}{\ra}}

\newcommand{\ptraceTanja}[1][k]{\pi_0(x_0) \prod_{i=1}^k \frac{w(x_{i-1},l_i,x_i)}{a(x_{i-1})}
\left (e^{-a(x_{i-1})\cdot \inf(I_i)}-e^{-a(x_{i-1})\cdot \sup(I_i)}\right)}

\def\longrightharpoonup{\relbar\joinrel\rightharpoonup}
\def\longleftharpoondown{\leftharpoondown\joinrel\relbar}

\def\longrightleftharpoons{
  \mathop{
    \vcenter{
      \hbox{
    \ooalign{
      \raise1pt\hbox{$\longrightharpoonup\joinrel$}\crcr
      \lower1pt\hbox{$\longleftharpoondown\joinrel$}
    }
      }
    }
  }
}

\newcommand{\agents}[2]{\funstyle{agents}\funtwo{#1}{#2}}
\newcommand{\card}[1][]{\funstyle{card}\fun{#1}}
\newcommand{\dom}[1][]{\funstyle{dom}\fun{#1}}

\newenvironment{myexample}{\begin{example}}{

\qed \end{example}}

\newcommand{\emptystate}{\epsilon}
\newcommand{\emptybinding}{\epsilon}
\newcommand{\emptylist}{\varepsilon}
\newcommand{\emptysite}{\epsilon}
\newcommand{\emptyagent}{\emptyset}

\newcommand{\agentname}{\mathcal {A}}
\newcommand{\sitename}{\mathcal {S}}
\newcommand{\internallabel}{\mathbb{I}}

\newcommand{\trans}[1][]{%
\def\argI{#1}%
\transrelay
}

\newcommand{\transrelay}[4][]{#2 \overset{#3%
{\def\arga{}
\ifx\arga\argI
\else%
{,\argI}%
\fi}%
}{\transarrow[#1]}#4}

\def\RR{\mathbb R}

\newcommand{\transarrow}[1][]{%
\def\arga{#1}%
\def\argb{\sharp}%
\ifx\arga\argb
\rightsquigarrow
\else%
\rightarrow_{#1}
\fi}

\begin{abstract}
When designing genetic circuits, the typical primitives used in major existing modelling formalisms 
are gene interaction graphs, where edges between genes denote either an activation or inhibition relation. 
However, when designing experiments, it is important to be precise about the low-level mechanistic details as to how each such relation is implemented. 
The rule-based modelling language Kappa allows to unambiguously specify mechanistic details such as DNA binding sites, dimerisation of transcription factors, or co-operative interactions. However, such a detailed description comes with complexity and computationally costly execution. We propose a general method for automatically transforming a rule-based program, by eliminating intermediate species and adjusting the rate constants accordingly. Our method consists of searching for those interaction patterns known to be amenable to equilibrium approximations (e.g.  Michaelis-Menten scheme). 
The reduced model is efficiently obtained by static inspection over the rule-set, and it represents a particular theoretical limit of the original model. 
The Bhattacharyya distance is proposed as a metric to estimate the reduction error for a given observable.
The tool is tested on a detailed rule-based model of a $\lambda$-phage switch, which lists $96$ rules and $16$ agents. 
The reduced model has $11$ rules and $5$ agents, and provides a dramatic reduction in simulation time of several orders of magnitude.
\end{abstract}




\section{Introduction}
\label{introduction}
One of the main goals of synthetic biology is to design and control genetic circuits in an analogous way to how electronic circuits are manipulated in human made computer systems. The field has demonstrated success in engineering simple genetic circuits that are encoded in DNA and perform their function in the cellular environment \cite{gardner2000construction}, \cite{guet2002combinatorial}. However, there remains a need for rigorous quantitative characterisation of such small circuits and their mutual compatibility \cite{kwok2010five}, which in electronic circuits is easily guaranteed by impedance matching. The important ingredient towards such characterisation is having an appropriate language for capturing model requirements, for prototyping the circuits, and for predicting their quantitative behaviour before committing to the time-intensive experimental implementation. 

Quantitative modelling of biomolecular systems is particularly challenging, because one deals with stochastic, highly dimensional, non-linear dynamical systems. For these reasons, modellers often immediately apply ad-hoc simplifications which neglect the mechanistic details, but allow to predict (simulate) the system's behaviour as a function of time. 
For example, the fact that \emph{protein $A$ activates protein $P$} is often modelled immediately in terms of a reaction $A\ra A+P$ with the Hill kinetic coefficient (e.g. $\frac{k[A]^n}{1+k[A]^n}$), while the mechanism in fact includes the formation of a macromolecular complex and its binding to a molecular target.
While such models are easier to execute, 
the simplification makes models hard to edit or refine. For example - a new experimental insight about an interaction mechanism cannot be easily integrated properly into the model, since several mechanistic steps are merged into a single kinetic rate.
Moreover, an abstract model does not provide precise enough design guide for circuit synthesis, and sometimes, only the more detailed models explain certain behaviours (e.g., in \cite{del2007design}, it is shown that only when incorporating the mRNA, the model explains certain experimentally observed facts).

Rule-based languages, such as Kappa \cite{webKappa} or BioNetGen \cite{webBNGL}, are designed to naturally capture the protein-centric and concurrent nature of biochemical signalling:
 the internal protein structure is maintained in form of a {\it site-graph}, and interactions can take place upon testing only {\it patterns}, local contexts of molecular species. 
 A site-graph is a graph where each node contains different types of sites, and edges can emerge from these sites. Nodes typically encode proteins and their sites are the protein binding-domains or modifiable residues; the edges indicate bonds between proteins. Then, every species is a connected site-graph, and a reaction mixture is a multi-set of connected site-graphs. 
The executions of rule-based models are traces of a continuous-time Markov chain (CTMC), defined according to the principles of chemical kinetics. 
In general, rule-based models are advantageous to the classical reaction models (Petri nets) for two major reasons. First, the explicit graphical representation of molecular complexes makes models easy to read, write, edit or compose (by simply merging two collections of rules). 
For example, the reaction of dimerization between two lambda $\agenttextfont{CI}$ molecules is classically written $2 CI \ra CI2$, where the convention is that $CI$ represents a free monomer, and $CI2$ represents a free dimer.  On the other hand, the same reaction written in Kappa amounts to:
\begin{align*}
\textit{'CI2:'}\;\;\;  \agent{CI}{ci,or}, \agent{CI}{ci,or} \leftrightarrow \agent{CI}{ci!1,or}, \agent{CI}{ci!1,or} @  \rate{2+},\rate{2-},
\end{align*}
where the binding sites $\agenttextfont{ci}$ and $\agenttextfont{or}$ are binding sites of the protein $\agenttextfont{CI}$, and $\agent{CI}{ci!1,or}$ denotes that the identifier of the rule-based bond account for the physical interaction between the two $\agenttextfont{CI}$ monomers, is $1$.
Secondly, a rule set can be executed, or subjected to formal static analysis: for example, it provides efficient simulations \cite{danos:aplas07}, \cite{hogg2014exact} automated answers about the reachability of a particular molecular complex \cite{VMCAI08}, or about causal relations between rule executions \cite{danos2007rule}.

The downside of incorporating too many mechanistic details in the model, is that they lead to computationally costly execution.
For this reason, we define and implement an efficient method for automatically detecting and applying equilibrium approximations. 
As a result, one obtains a smaller model, where some species are eliminated, and the kinetic rates are appropriately adjusted.
In this way, the experimentalist can choose to obtain the predictions more efficiently but less accurately, however without losing track of the underlying low-level mechanisms. 

To the best of our knowledge, there exist no efficient methods to quantify the error induced by time-scale separation approximations for biochemical reaction networks.
The bottleneck is the complexity of the original system, whose behaviour is computationally costly to analyse - often even to run a single simulation trace.
The correctness of our approach relies on the fact that the approximate model is equal to the original one, in the artificial limit where certain reactions happen at a sufficiently larger time-scale than others, and they are seemingly equilibrated shortly upon the reactions initiate. 
Faced with designing or modelling biological circuits with many connections and highly heterogeneous hardware, the ability of predicting solutions that lack a precise error measurement is of secondary importance. What is desirable is to have a prototype or model of a circuit that displays the desired behaviour at a qualitative level, which later on can be further improved. Furthermore, most kinetic rates are rarely precisely determined experimentally, and hence precise quantitative error estimates are not necessarily relevant and on top are time consuming, when faced with imprecise input characteristics of the underlying prototype model.

\textbf{Implementation and testing}. 
The tool is implemented in OCaml, and it is tested on a detailed rule-based model of a $\lambda$-phage switch \cite{ptashne1986genetic}, \cite{ptashne2001genes}. 
Simulations were carried out on 
the complete chemical reaction genetic circuit model which contains 96 rules, 16 agents and 61 species. 
The  model is reduced to only 11 rules and 5 agents. 
It is worthwhile emphasising that our reduction method is general -- applicable to any rule-set, and that the reduced model is obtained almost instantaneously.

\textbf{Related work.}
Our method is inspired from the work presented in \cite{myers2011engineering} and \cite{kuwahara2006automated}, with the adaptations which arise due to the differences between  the reaction lists and the rule-based language.
Apart from rule-based models, other formalisms were proposed for characterisation of synthetic devices, such as, e.g., linear temporal logic \cite{bartocci2013temporal}. 
In a broader context, the principle of obtaining conclusions about system's dynamics by analyzing their model description, originates from, and is exhaustively studied in the field of formal program verification and model checking \cite{CousotAI}, \cite{burch1992symbolic}, while it is recently gaining recognition in the context of programs used for modeling biochemical networks. An example is the related work of detecting fragments for reducing the deterministic or stochastic rule-based models \cite{pnas2009}, \cite{mathjournal}, \cite{lumpabilityTCS}, detecting the information flow for ODE models of biochemical signaling \cite{conzelmann,borisov2005_1}, or the reaction network theory \cite{craciun2006multiple}. Related works on systematic model reduction techniques are based on separating time-scales  \cite{hepp2014adaptive,rao2003stochastic,kang2013separation,ovidiu,hogg2014exact}, or they propose numerical algorithms which focus of efficiently obtaining the evolution of the probability distribution over time (the master equation) \cite{munsky2006finite,VerenaSW}.

\textbf{Paper outline.}
\secref{preliminaries} introduces two concepts: 
(i) the classical stochastic and deterministic model of chemical reaction networks, and 
(ii) the rule-based modelling language Kappa.
\secref{approximation} illustrates the equilibrium approximation schemes (the generalized Michaelis-Menten and fast dimerisation) and discusses the theoretical guarantees about the approximate system.
In \secref{algorithms}, we outline the algorithms for detecting the approximation schemes (operator site reduction and dimerisation reduction).
Finally, in \secref{results}, we describe the $\lambda$-phage model and we compare the results and the CPU time for the original and approximate model.

\section{Preliminaries}
\label{preliminaries}
\begin{figure}[t]
\begin{center}
\includegraphics[scale=0.4]{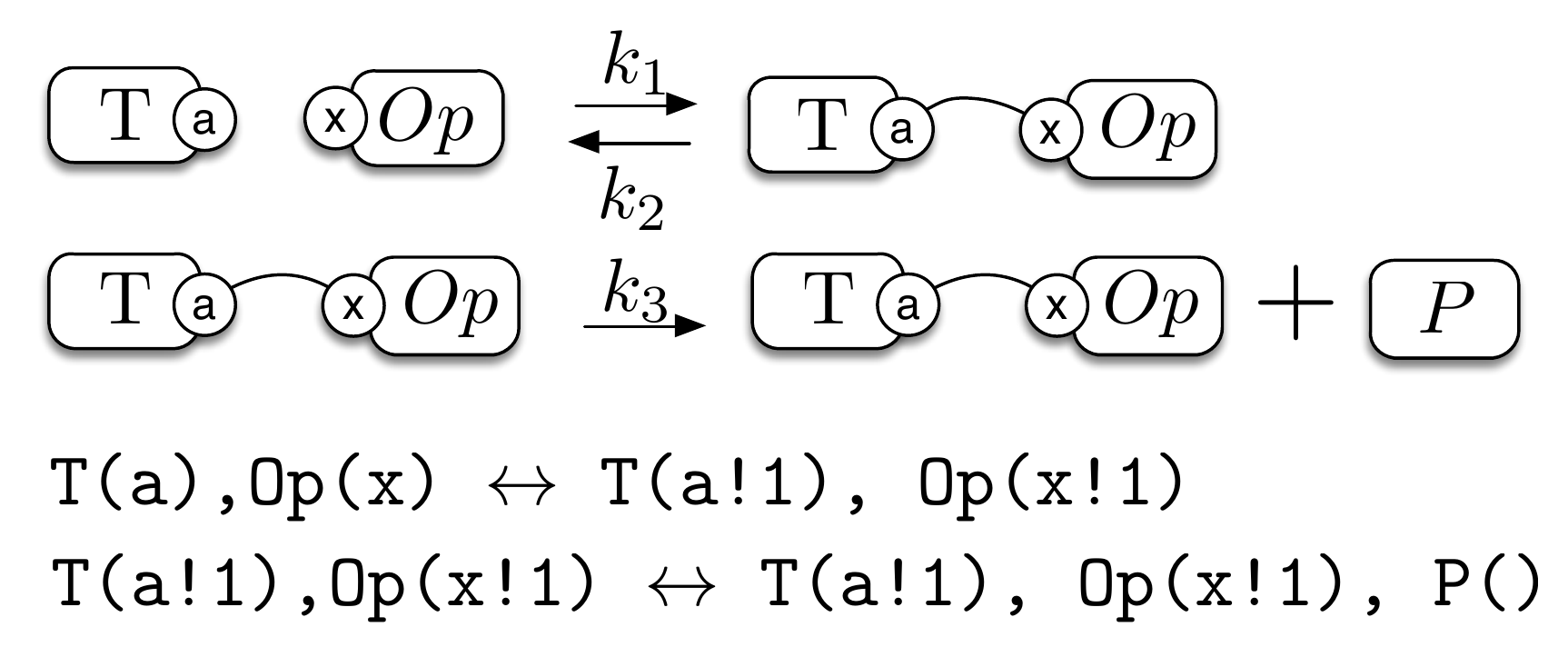}
\end{center}
\caption{
An example of a rule-based model. The transcription factor $T$ binds to the operator and initiates the production of protein $P$.
}
\label{fig:rulebased}
\end{figure}

\subsection{Stochastic chemical reaction networks}
For a well-mixed reaction system with molecular species $\speciesSet=\{\spec{1},\ldots,\spec{\specN}\}$, the state of a system can be represented as a multi set of those species, denoted by 
$\xstov=(\xsto_1,...,\xsto_{\specN})\in \N^{\specN}$. 
The dynamics of such as system is determined by a set of reactions $\reactionSet=\{\reac_1,\ldots,\reac_{\reacN}\}$. 
Each reaction is a triple 
$\reac_j\equiv(\consv{j},\changev_{j},\srate_j)\in \N^{\specN}\times\N^{\specN}\times \Rpos$, written down in the following form: 
\begin{align*}
\cons{1j}\spec{1},\ldots,\cons{\specN j}\spec{\specN} \rA{\drate_{j}}\prdn{1j}\spec{1},\ldots,\prdn{\specN j}\spec{\specN}, \\
 \hbox{ such that $\prdn{ij}=\cons{ij}+\change_{ij}$}.
\end{align*}
The vectors $\consv{j}$ and $\consv{j}'$ are often called respectively the \emph{consumption} and \emph{production} vectors due to reaction $\reac_j$, and $k_j$ is the \emph{kinetic rate} of reaction $\reac_j$.
If the reaction $\reac_j$ occurs, after being in state $\xstov$, the next state will be 
$\xstov' = \xstov+\changev_j$.
This will be possible only if $\xsto_i\geq \cons{ji}$ for all $i=1,\ldots,\specN$.
 Under certain physical assumptions \cite{Gillespie}, the species multiplicities follow a continuous-time Markov chain (CTMC)  $\{\procX(t)\}$, defined over the state space $\states=\{\xstov\mid \hbox{$\xvec$ is reachable from $\xstov_0$ in $\reactionSet$}\}$.
 Hence, the probability of moving to the state $\xstov+\changev_j$ from $\xstov$ after time $\Delta$ is
 \[
 \PP(\procX(t+\Delta) = \xstov+\change_k\mid \procX(t)=\xstov) = \sprop{j}(\xstov)\Delta+o(\Delta),
 \]
 with $\sprop{j}$ the propensity of $j$th reaction, assumed to follow the principle of mass-action: 
 $\sprop{j}(\xstov)= 
\drate_{j}\prod_{i=1}^{\specN}{x_i\choose\cons{ij}}
$. 
The binomial coefficient ${x_i\choose\cons{ij}}$ reflects the probability of choosing $\cons{ij}$ molecules of species $S_i$ out of $x_i$ available ones.

\subsection{Deterministic limit} 
In the continuous, deterministic model of a chemical reaction network, the state $\zvec(t) = (z_1,\ldots,z_n)(t)\in\R^n$ is represented by listing the concentrations of each species.
The dynamics is given by a set of differential equations in form
\begin{align}
\label{eq:ode}
\deriv z_i = \nu_{ij}\sum_{j=1}^\reacN c_j \prod_{i=1}^nz_i(t)^{a_{ij}},
\end{align}
where $c_j$ is a deterministic rate constant, computed from the stochastic one and the volume $\volume$ from $c_j = k_j\volume^{|\consv{j}|-1}$ ($|\mathbf{x}|$ denotes the $1$-norm of the vector $\mathbf{x}$).
The deterministic model is a limit of the stochastic model when all species in a reaction network are highly abundant. 
Denote by $\Rc{j}(t)$ the number of times that the $j$-th reaction had happened until the time $t$.
Then, the state of the stochastic model at time $t$ is  
\begin{align}
\label{eq:rtc}
\Xstov(t) = \Xstov(0) + \sum_{j=1}^{\reacN} \Rc{j}(t)\changev_{j}.
\end{align}
The value of $\Rc{j}(t)$ is a random variable, that can be described by a non-homogenous Poisson process, with parameter $\int_0^t \sprop{j}(\Xvec({s}))\der s$, that is, 
$\Rc{j}(t) = \pois_j(\int_0^t \sprop{j}(\Xstov({s}))\der s)$.
Then, the evolution of the state $\Xstov(t)$ is given by the expression
\begin{align}
\Xstov(t) = \Xstov(0) + \sum_{j=1}^{\reacN}  \pois_j\left(\int_0^t \sprop{j}( \Xstov(s){})\der s\right) \gainv{j}
\label{SSKpp}.
\end{align}

By scaling the species multiplicities with the volume: $Z_i(t) = X_i(t)/\volume$, adjusting the propensities accordingly, 
 in the limit of infinite volume $\volume\ra \infty$, the scaled process ${\mathbf Z}(t)$ follows an ordinary differential equation \eqref{eq:ode} \cite{kurtztg_1971_1}.

%
It is worth mentioning here that the above scaling from stochastic to the deterministic model is a special case of a more general framework presented in \cite{kang2013separation}, referred to as the \emph{multiscale stochastic reaction networks}.
Intuitively, the deterministic model is a special case where all species are scaled to concentrations and reaction rates are scaled always in the same way, depending on their arity. 
The reductions shown in this paper will be based on a variant of multiscale framework, where some species are scaled to concentrations and others are kept in copy numbers, and where reaction rates have varying scales as well.

%

%
%
%

\subsection{Rule-based Models}

In this section, we introduce the rule-based modeling language Kappa,
which is used to specify chemical reaction networks, by explicitly denoting chemical species in form of site-graphs. 
A simple example of a Kappa model is presented in \figref{fig:rulebased}.

For the stochastic semantics of Kappa, that is a continuous-time Markov chain (CTMC) assigned to a rule-based model, we refer to \cite{conf/vmcai/DanosFFK08} or \cite{lumpabilityTCS}.
Intuitively, any rule-based system can be expanded to an equivalent reaction system (with potentially infinitely many species and reactions).
The stochastic semantics of a Kappa system is then the CTMC $\{X(t)\}$ assigned to that equivalent reaction system. 
Even though the semantics of a Kappa system is defined as the semantics of the equivalent reaction system, in practice, using Kappa models can be advantageous for several reasons -
they are easy to read, write, edit or compose, they can compactly represent potentially infinite set of reactions or species, and, most importantly, they can be symbolically executed.

We present Kappa in a process-like notation. 
We start with an operational semantics. 

Given a set $X$, $\wp(X)$ denotes the power set of $X$ (i.e.~the set of all subsets of $X$).
We assume a finite set of agent names $\agentname$, representing different kinds of proteins; a finite set of sites $\sitename$, corresponding to protein domains; a finite set of internal states $\stateset$, and $\interface_{\iota}$, $\interface_{\beta}$ two signature maps from $\agentname$ to $\wp(\sitename)$, listing the domains of a protein which can bear respectively an internal state and a binding state. We denote by $\interface$ the signature map that associates to each agent name $A\in\agentname$ the combined interface $\interface_\iota(A)\cup\interface_\beta(A)$.

\newdef{definition}{Definition}
\begin{definition} (Kappa agent) \rm
A \emph{Kappa agent} $A(\sigma)$ is defined by its type $A\in \agentname$ and  its \emph{interface} $\sigma$. In $A(\sigma)$, the interface $\sigma$ is a sequence of sites $s$ in $\interface(A)$, with internal states (as subscript) and binding states (as superscript). The internal state of the site $s$ may be written as $\site{s}_{\epsilon}$, which means that either it does not have internal states (when $s\in \interface(A)\setminus  \interface_{\iota}(A)$), 
or it is not specified. A site that bears an internal state $m\in\stateset$ is written $\site{s}_{m}{}$ (in such a case $s\in\interface_\iota(A)$). The binding state of a site $s$ can be specified as $\site{s}^{\epsilon}$, if it is \emph{free}, otherwise it is bound (which is possible only when $s\in\interface_\beta(A)$). There are several levels of information about the binding partner: we use a binding label $i\in\bindinglabel$ when we know the binding partner, or a wildcard bond $\bound{}$ when we only know that the site is bound. The detailed description of the syntax of a Kappa agent is given by the following grammar: 
\begin{equation*}
\begin{array}{rcll}
a & ::= & \agent{$N$}{$\sigma$} & \hbox{(agent)}\cr
\agent{$N$}{} & ::= & A\in \agentname & \hbox{(agent name)} \cr
\sigma & ::= & \emptylist \;\mid\; s{\sitesep}\sigma & \hbox{(interface)}\cr
s  & ::=  & {\site{n}_{\iota}^{\lambda}} & \hbox{(site)}\cr
{\site{n}{}{}} & ::=  & x\in\sitename & \hbox{(site name)} \cr
\iota & ::= & \emptystate \;\mid\; m\in\stateset & \hbox{(internal state)}\cr
\lambda & ::= & \emptybinding \;\mid\; \bound{}\;\mid\;i\in \bindinglabel & \hbox{(binding state)} \cr
\end{array}
\end{equation*}
We generally omit the symbol $\emptybinding$.
\end{definition}

\begin{definition} (Kappa expression) \rm
\emph{Kappa expression} $E$ is a set of agents \agent{A}{$\sigma$} and fictitious  agents $\emptyagent$. Thus the syntax of a Kappa expression is defined as follows:
\begin{align*}
E & ::= \emptylist \;\mid\; a{\agentsep}E \;\mid\; \emptyagent{\agentsep}E.
\end{align*}
\end{definition}

The structural equivalence $\equiv$, defined as the smallest binary equivalence relation between expressions that satisfies the rules given as follows
\begin{align*}
\begin{array}{c}
 \begin{array}{rcl}
   E{\agentsep}\agent{$A$}{$\sigma{\sitesep}s{\sitesep}s'{\sitesep}\sigma'$}{\agentsep}E' & \equiv & E{\agentsep}\agent{$A$}{$\sigma{\sitesep}s'{\sitesep}s{\sitesep}\sigma'$}{\agentsep}E' \\
   E{\agentsep}a{\agentsep}a'{\agentsep}E' & \equiv & E{\agentsep}a'{\agentsep}a{\agentsep}E' \\
   E & \equiv & E{\agentsep}\emptyagent \end{array} \cr
\begin{array}{c}
\strutfrac{\displaystyle
  i,j \in \bindinglabel \textnormal{ and }
  i \textnormal{ does not occur in } E}{
  \displaystyle E[i/j] \equiv E} \\
\strutfrac{\displaystyle 
  i \in \bindinglabel \textnormal{ and } i \textnormal{ occurs only once in } E}{  \displaystyle E[\emptybinding/i] \equiv E}
\end{array}
\end{array}
\end{align*}
stipulates that neither the order of sites in interfaces nor the order of agents in expressions matters, that a fictitious agent might as well not be there, that binding labels can be injectively renamed and that \emph{dangling bonds} can be removed.


\begin{definition} (Kappa pattern,mixture and species)\\ \rm
A {\em Kappa pattern} is a Kappa expression which satisfies the following five conditions: (i) no site name occurs more than once in a given interface; (ii) each site name $s$ in the interface of the agent $A$ occurs in $\interface(A)$; (iii) each site $s$ which occurs in the interface of the agent $A$ with a non empty internal state occurs in $\Sigma_\iota(A)$; (iv) each site $s$ which occurs in the interface of the agent $A$ with a non empty binding state occurs in $\Sigma_\lambda(A)$; and (v) each binding label $i\in\bindinglabel$ occurs exactly twice if it does at all --- there are no dangling bonds. A \emph{mixture} is a pattern that is fully specified, i.e.~each agent $A$ documents its full interface $\interface(A)$, a site can only be free or tagged with a binding label $i\in\bindinglabel$, a site in $\interface_\iota(A)$ bears an internal state in $\stateset$, and no fictitious agent occurs.
A \emph{species} is a connected mixture, i.e.~for each two agents $A_0$ and $A$ there is a finite sequence of agents $A_1,\ldots,A_k$ s.t. there is a bond between a site of $A_k$ and of $A$ and  for $i=0,1,\ldots,k-1$, there is a site of agent $A_i$ and a site of agent $A_{i+1}$.
\end{definition}

\begin{definition} (species occurring in a pattern) \rm
Given Kappa patterns $E_s$ and $E_p$, if $E_s$ defines a Kappa species, and $E_s$ is a substring of $E_p$, we say that a species $E_s$ \emph{occurs} in a pattern $E_p$.
\end{definition}

\begin{definition} (Kappa rule) \rm
A Kappa rule $r$ is defined by two Kappa patterns $E_\ell$ and $E_r$, and a rate $k\in \Rpos$, and is written: $r = {E_\ell \rightarrow E_r @ k}$.

A rule $r$ is well-defined, if the expression $E_r$ is obtained from $E_\ell$ by finite application of the following operations: (i) creation (some fictitious agents $\emptyagent$ are replaced with some fully defined agents of the form \agent{A}{$\sigma$}, moreover $\sigma$ documents all the sites occurring in $\interface(A)$ and all site in $\interface_\iota(A)$ bears an internal state in $\stateset$), (ii) unbinding (some occurrences of the wild card and binding labels are removed), (iii) deletion (some agents with only free sites are replaced with fictitious agent $\emptyagent$), (iv) modification (some non-empty internal states are replaced with some non-empty internal states), (v) binding (some free sites are bound pair-wise by using binding labels in $\bindinglabel$). 
\end{definition}

In our static inspection of rules, we will test species (fully defined connected mixtures). 
To this end, we adopt the terminology of \emph{reactant}, \emph{modifier} and \emph{product} from \cite{kuwahara2006automated}.

\begin{definition} (reactant, modifier, product) \rm
Given a rule $(E_l,E_r)$, a Kappa species $\specP$ is called 
\begin{itemize}
\item a \emph{reactant}, if it occurs in pattern $E_l$ and does not occur in pattern $E_r$ , 
\item a \emph{modifier}, if the number of occurrences in pattern $E_l$ equals the number of occurrences in pattern $E_r$ , 
\item a \emph{product}, if it does not occur in pattern $E_l$, and it occurs in pattern $E_r$.
\end{itemize}
\end{definition}

\begin{definition} (Kappa system) \rm
A \emph{Kappa system} $\mathcal{R} (\xvec_0,\observs,$ 
$\{r_1,\ldots,r_n\})$ is given by an initial mixture $\xvec_0$, a set of Kappa patterns $\observs$ called {\em observables}, and a finite set of rules $\{r_1,\ldots,r_n\}$. 
\end{definition}


\section{Model Approximation}
\label{approximation}

\begin{figure}[t]
\begin{center}
\includegraphics[scale=0.13]{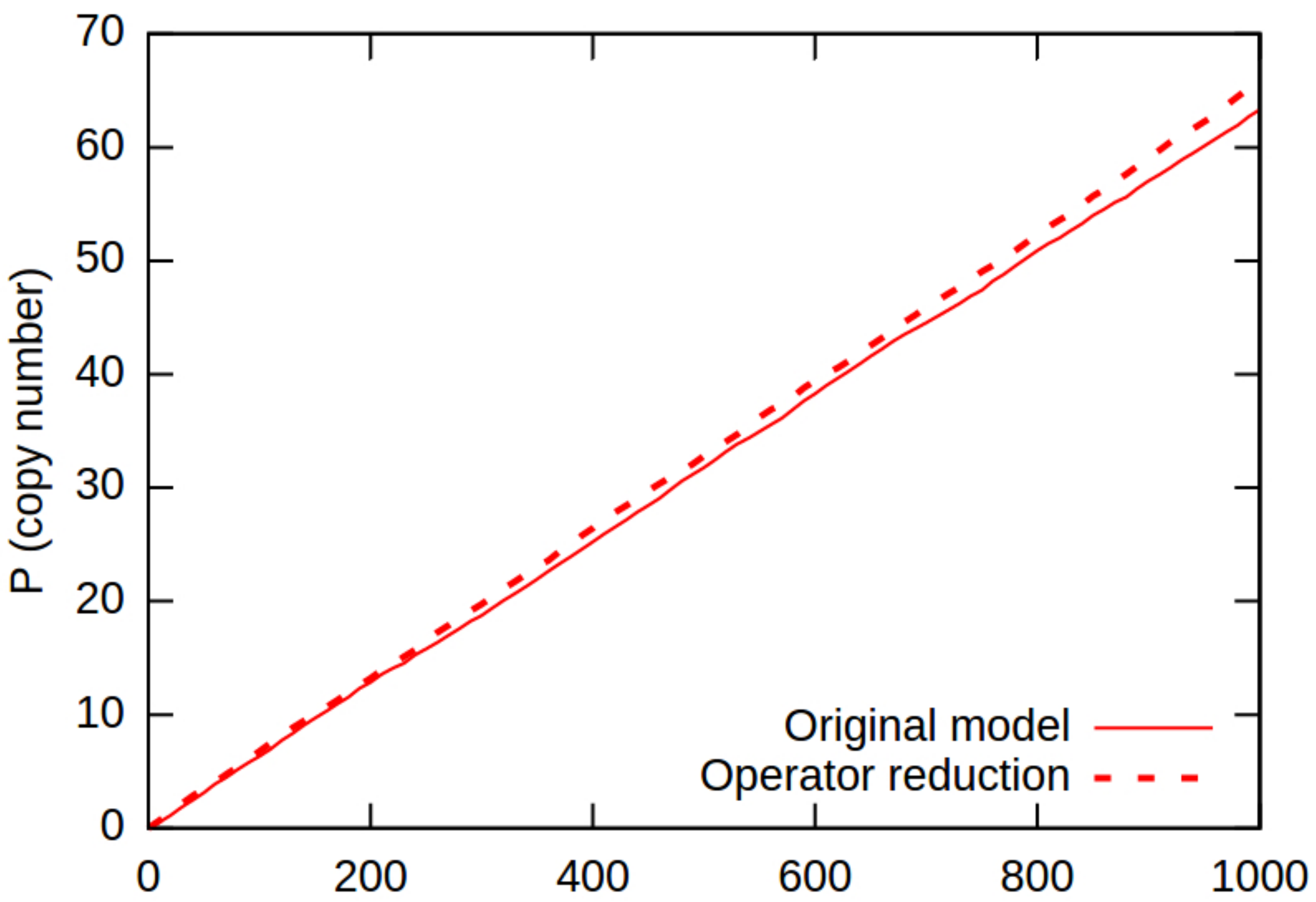}
\includegraphics[scale=0.13]{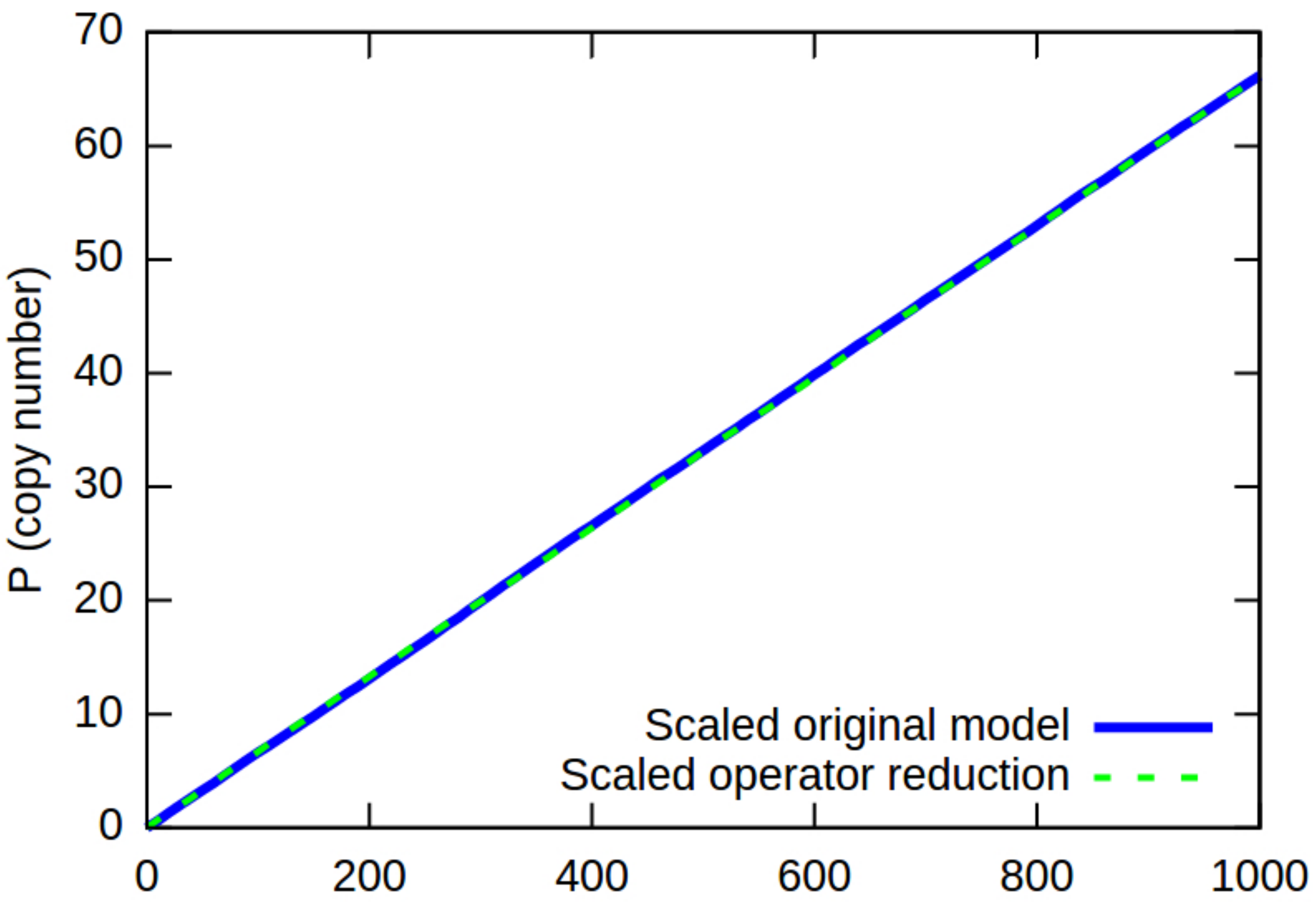}\\
a) \hspace{23mm}
b)\\
\includegraphics[scale=0.15]{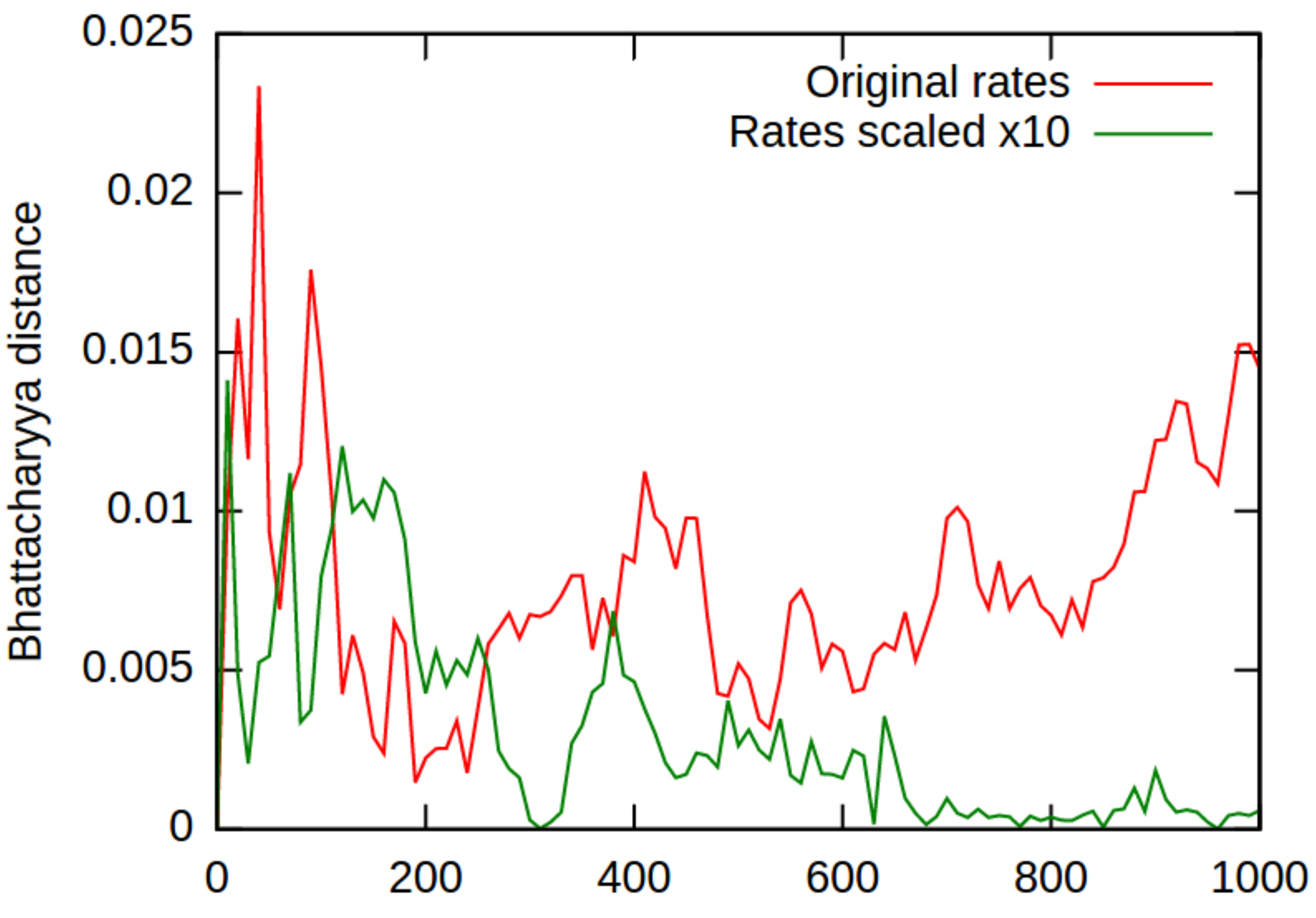}\\
\hspace{5mm} c)
\end{center}
\caption{
Example shown in \figref{fig:rulebased}.
The mean protein expression for one hundred sampled traces, before and after the \mm\;reduction.
a) Parameters $k_1=0.2156$, $k_2=1$, $k_3=0.014$ and there are initially $50$ transcription factors.
The mean and standard deviation (not shown) are computed for each time point, for the original (full line) and reduced model (dotted line).
%
b) Parameters $k_2$, $k_3$, and the initial number of transcription factors $T$ are scaled up by factor $\volume = 10$. Same notation as for a)
%
c) 
The Bhattacharyya distance between the distributions of the protein level with a model before and after the reduction.
Red plot refers to the parameter values shown in a), and the green plot to the scaled parameter values shown in b).
%
%
}
\label{fig:exampleMM}
\end{figure}

In this section, we present the mathematical analysis underlying the approximation algorithms presented in \secref{algorithms}.
Our reductions will be based on three reduction schemes: \mm\;reduction, \mmgen\;reduction and \fastdim\;reduction.

\subsection{Enzymatic reduction}
\label{sec:\MM}
%
Assume the elementary enzymatic transformation from a substrate $\subs$ to a product $\product$, through the intermediate complex $E:S$:
\begin{align}
\label{eq:mmreac}
\enz+\subs\rates{k_1}{k_2} \enz:\subs \rA{{k_3}} \enz+\product,
\end{align}
which our algorithm will convert to the well-known Michaelis-Menten form
\begin{align}	
\label{eq:mm}
\subs\rA{\frac{k_3 \enz_T K}{1+K\conc{S}}} \product,
\end{align}
where $\enz_T=\conc{E}(t)+\conc{E:S}(t)$ denotes the total concentration of the enzyme, 
and $K=\frac{k_1}{k_2+{k_3}}$.  

The above approximation is generally considered to be sufficiently good under different assumptions, 
such as, for example, that the rate of dissociation of the complex to the substrate is much faster than its dissociation to the product (i.e. $k_2\gg k_3$), 
also known as the \emph{equilibrium approximation}.  
Even if the equilibrium condition is not satisfied, it can be compensated in a situation where the total number of substrates significantly outnumbers the enzyme concentration - ${\conc{S}(0)+K}\gg {\enz_T}$, known as the \emph{quasi-steady-state assumption}.


Whenever one of the above assumptions holds, the quantity of the intermediate complex can be assumed to be rapidly reaching equilibrium, that is, $\deriv \conc{E:T}(t) =0$.
Then, it is straightforward to derive the rate of direct conversion from substrate to product:
\begin{align*}
\deriv\conc{\product} = \frac{k_3\enz_TK}{1+K\conc{S}}\conc{S},
\end{align*}
which exactly corresponds to the equation for the rule \eqref{eq:mm}.

The informal terminology of being `significantly faster', motivated the rigorous study of the limitations of the approximations based on separating time scales. 
While the enzymatic (Michaelis-Menten) approximation has been first introduced and subsequently studied in the context of deterministic models (e.g. \cite{murray2002mathematical}, Ch.6), it was more recently that the time-scale separation was investigated in the stochastic context \cite{rao2003stochastic}, \cite{haseltine2002approximate}, \cite{crudu2009hybrid}, \cite{hepp2014adaptive}, \cite{sanft2011legitimacy}, \cite{gillespie2009subtle}.
%
Notably, the following result from \cite{darden1979pseudo} (also shown as a special case of the multi scale stochastic analysis from  \cite{kang2013separation}), shows that,
under an appropriate scaling of species' abundance and reaction rates,
 the original model and the approximate model converge to the same process.

\newtheorem{theorem}{Theorem} 
\begin{theorem} (Darden \cite{darden1979pseudo}, Kang \cite{kang2013separation}). \rm
Consider the reaction network \eqref{eq:mmreac} (equivalently the rule-based system depicted in \figref{fig:exampleMM}), 
and denote by $X_{\subs}(t)$, $X_{\enz}(t)$, $X_{\enz:\subs}(t)$ and $X_{\product}(t)$ the 
copy numbers of the respective species due to the random-time change model \eqref{eq:rtc}.
Denote by $\enz_T = X_{\enz:\subs}(t)+X_{\enz}(t)$ and
$V_{\enz}(t) = \int_0^t \volume^{-1}X_{\enz}(s)\der s$
 and assume that $\volume = {\cal O}(X_{\subs})$. 
Assume that
$k_1\ra\gamma_1$,
$k_2/\volume \ra\gamma_2$,
$k_3/\volume \ra \gamma_3$,
$\volume\to\infty$, and
$\frac{X_{\subs}(0)}{\volume}\ra x_{\subs}(0)$.
Then $(\frac{X_{\subs}(t)}{\volume}, \Vel_{\enz}(t))$ converges to $(\conc{\subs}(t), \vel_{\enz}(t))$ and 
\[\deriv\vel_{\enz}(s) = \frac{\enz_T}{1+\hat{K}\conc{\enz}(s)}
\hbox{ \;\;\; and \;\;\;} 
\deriv\conc{\subs} = - \frac{\enz_T{\gamma_3} \hat{K}\conc{\subs}(t)}{1+\hat{K}\conc{\subs}(t)},
\]
where $\hat{K} = \frac{\gamma_1}{\gamma_2+{\gamma_3}}$.
\label{thm:kang}
\end{theorem}

The assumptions listed in the theorem capture the that:
(i) $X_{\subs}$ and $X_{\product}$ are scaled to concentrations, while $X_{\enz}$ and $X_{\enz:\subs}$ remain in copy numbers;
(ii) the stochastic reaction rate $k_1$ is an order of magnitude smaller than the rates $k_2$ and $k_3$ (as a consequence of being related to the bimolecular, and not unimolecular reaction).

A complete proof is provided in \cite{kang2013separation}. 
We here outline the general idea.
Let $N>0$ be a natural number, and let $Z_{\subs}(t) = X_{\subs}(t)/\volume$, $Z_{\enz}(t) = X_{\enz}(t)$, $Z_{\subs:\enz}(t) = X_{\subs:\enz}(t)$, $Z_{\product}(t) = X_{\product}(t)/\volume$.
Writing out the scaled random time-change model for the substrate gives:

\begin{align*}
Z_{\subs}(t)  = Z_{\subs}(0) &- \volume^{-1}\pois_1(\volume \int_{0}^t \gamma_1 Z_{\subs}(s)Z_{\enz}(s)\der s)\\
&+\volume^{-1}\pois_2(\volume \int_{0}^t \gamma_2 Z_{\subs:\enz}(s)\der s),
\end{align*}

and writing out the scaled random time-change model for the complex gives:

\begin{align*}
Z_{\enz:\subs}(t) = Z_{\enz:\subs}(0) & + \pois_1(\volume \int_{0}^t \gamma_1 Z_{\subs}(s)Z_{\enz}(s)\der s)\\
& -\pois_2(\volume \int_{0}^t \gamma_2 Z_{\subs:\enz}(s)\der s)\\
& -\pois_3(\volume \int_{0}^t \gamma_3 Z_{\subs:\enz}(s)\der s).
\end{align*}

After dividing the latter with $\volume$, and applying the law of large numbers, we obtain the balance equations analogous to assuming that the complex is at equilibrium.
This equation implies the expression for $\frac{\der}{\der t} v_{\enz}(s)$.
The equation for $\deriv\conc{\subs}$ follows from the model of $Z_{\subs}(t)$:
we first use the conservation law $Z_{\subs:\enz}(s)=\volume^{-1}\enz_{T}-Z_{\enz}(t)$ and then substitute the obtained value of $\frac{\der}{\der t} v_{\enz}(s)$.
%
%

In order to confirm that the reduction is appropriate, our goal is now to show that the scaled versions of the original model \eqref{eq:mmreac} and the reduced model \eqref{eq:mm} are equivalent in the limit when $\volume\ra\infty$.
Let $Z_{\product}(t) := \volume^{-1} X_{\product}(t)$ be the scaled random time change for the product in the original model, and $\hat{Z}_{\product}(t):=\volume^{-1}\hat{X}_{\product}(t)$ in the reduced model. 
Notice that, from the balance equations, $\deriv{x_{\product}}=-\deriv{x_{\subs}}$.
According to the reduced system \eqref{eq:mm}, the random time change for the product is given by
\begin{align*}
\hat{Z}_{\product}(t) & = \hat{Z}_{\product}(0)+\volume^{-1}\pois(\int_0^t \frac{k_3\enz_{T} K}{1+K\volume \hat{Z}_{\subs}(s)}\volume\hat{Z}_{\subs}(s) \der s)\\
& = \hat{Z}_{\product}(0)+\volume^{-1}\pois(\int_0^t \volume\frac{\gamma_3\enz_{T} \hat{K}}{1+\hat{K} \hat{Z}_{\subs}(s)}\hat{Z}_{\subs}(s) \der s).
\end{align*}
Passing to the limit, we obtain the desired relation
$\deriv{\hat{z_{\product}}(t)}=\deriv{z_{\product}(t)}$.

The above Theorem does not provide the means of computing the approximation error, or an algorithm which suggests which difference in time-scales is good enough for an approximation to perform well.
Rather, this result shows 
that the enzymatic approximation is justified in the limit 
when the assumptions about the reaction rates and species' abundance are met. 
In other words, 
when $N\ra\infty$,
the scaled versions of the original and reduced models -- e.g. $Z_{\product}(t)=\volume^{-1}X_{\product}(t)$ and $\hat{Z}_{\product}=\volume^{-1}\hat{X}_{\product}$ -- both converge to at the same, well-behaved process. 
This provides confidence that the actual process $\hat{X}_{\product}$ is a good approximation of the process $X_{\product}$.
%

In our reduction algorithm (\secref{algorithms}), we will apply the reduction whenever the pattern \eqref{eq:mmreac} is detected.
In order to ensure the validity of the approximation in the context of other rules, we will additionally check that the enzyme $\enz$ (resp. complex $\enz:\subs$)
have initially low copy number (zero resp.), and that they don't appear in any other rule (unless it is another \mm\, scheme).

\begin{example}\rm
To illustrate the meaning of the \thmref{thm:kang}, we apply our reduction method on a small example shown in \figref{fig:rulebased}.
We plot the mean and we compute the standard deviation of the protein level for the original and for the reduced model.
Then, we scale up the parameters $k_2$ and $k_3$, as well as the initial concentration of transcription factor $T$, in order to mimic the effect of choosing a larger $N$ in \thmref{thm:kang}.
The deviation between the curves is decreased, as can be seen in \figref{fig:exampleMM}.
In order to obtain the error of using the reduced system instead of the original one, we compute the Bhattacharyya distance for each time point, for the actual parameter set and for the scaled parameter set.
%
As expected, the distance is overall smaller in the scaled system.
Especially in the scaled system (green line), we can observe that initially, the distance is larger, and then it decreases with time.
This is because the original system takes time to reach the equilibrium state which is, in the reduced system, assumed immediately.
\end{example}

\subsection{Generalised enzymatic reduction}

\label{sec:generalized}
The enzymatic approximation can be generalized to a situation where many sets of substrates compete for binding to the same enzyme.
Consider a sub-network of $n$ reactions where the $i$-th such reaction reads:
\begin{align*}
\enz+\subs_{i,1}+\ldots+\subs_{i,m_i} & \rates{k_i}{k_i^-} \enz:\subs_{i,1}:\ldots:\subs_{i,m_i} \rA{\hat{k_i}} \enz+\product_i. 
\end{align*}
The resulting approximation is
\begin{align*}
\subs_{i,1}+\ldots+\subs_{i,m_i}\rA{\frac{\hat{k_i}\enz_T K_1\conc{\subs_i}}{Z}} \product_i,
\end{align*}
where $\conc{\subs_i} = \prod_{j\in\{1,\ldots,m_i\}}\conc{\subs_{i,j}}$,  
$Z= 1+\sum_{j\in\{1,\ldots,n\}}\conc{\subs_j}$
and 
$\enz_T=\conc{E}(t)+\sum_{i=1}^n\conc{E:S_{i,1}:\ldots:S_{i,m_i}}(t)$. 
The latter expression follows from $\deriv \conc{E:S_{i,1}:\ldots:S_{i,m_i}}(t) =0$ for all $i=0,\ldots,n$.

The correctness of the generalized enzymatic reduction can be shown with the same technique as \thmref{thm:kang}.
Each substrate and product should be scaled to concentrations, while all intermediate complexes and the enzyme remain in copy numbers.
The relations between the reaction rates are equivalent.
%
%

\subsection{\Fastdim\;reduction}
\label{sec:dimerization}
%
Consider now the dimerisation reaction $M+M\rates{k}{k^-} M_2$.
Assuming that both rates $k$ and $k^-$ are fast comparing to other reactions involving $M$ or $M_2$, 
it is common to assume that the reaction is equilibrated, that is, 
$kx_M(t)^2-k^-x_{M2}(t)=0$, where $x_M(t)$ and $x_{M2}(t)$ denote the copy number at time $t$, of monomers and dimers respectively.
%
Such assumption allows us to eliminate the dimerization reactions,
and only the total amount of molecules $M$ needs to be tracked in the system.
%
%
The respective monomer and dimer concentrations can be expressed as fractions of the total concentration:
\begin{align*}
\conc{M}(t)  = \frac{1}{4K} \left(\sqrt{8K{M_T}(t)+1} -1 \right),\;\;\;\hbox{     and     } \;\;\; \\
\conc{M_2}(t)  = \frac{M_T(t)}{2}-\frac{1}{2}\conc{M}(t),
\end{align*}
where $K=\frac{k}{k^-}$ and $M_T(t) = \conc{M}(t)+2\conc{M_2}(t)$.

The correctness of the generalized enzymatic and dimerization reduction can be shown with the same technique as \thmref{thm:kang}.
In the context of multiscale stochastic reaction networks, both reaction rates should be treated as fast.

\section{Reduction Algorithm}
\label{algorithms}

\begin{algorithm}[t]
\SetCommentSty{emph}
\SetKwData{Left}{left}\SetKwData{This}{this}\SetKwData{Up}{up} \SetKwFunction{Union}  {Union}\SetKwFunction{FindCompress}{FindCompress} \SetKwInOut{Input}{Input}\SetKwInOut{Output}  {Output} 
\Input{
A Kappa system $\mathcal{R} = (\xvec_0,\observs,\{r_1,\ldots,r_n\})$. 
} 
\Output{
A Kappa system $\mathcal{R}' = (\xvec'_0,\observs,\{r_1',\ldots,r_m'\})$.
}
$\mathcal{R}  \longleftarrow$ ME($\mathcal{R} $), $\mathcal{R} \longleftarrow$ SRC($\mathcal{R} $)\\
$\mathcal{R} \longleftarrow$ Fast dimerization reduction($\mathcal{R} $)\\
$\mathcal{R}  \longleftarrow$ ME($\mathcal{R} $), $\mathcal{R} \longleftarrow$ SRC($\mathcal{R} $)\\
$\mathcal{R} \longleftarrow$ \MMgen\, reduction ($\mathcal{R} $)\\
$\mathcal{R}  \longleftarrow$ ME($\mathcal{R} $), $\mathcal{R} \longleftarrow$ SRC($\mathcal{R} $)\\
\caption{Approximation algorithm. `ME' is a shorthand for `modifier elimination' and `SRC' is shorthand for similar rule composition. 
The exact reductions are performed before and after each of the two other reductions.
%
} 
\label{algo:Red} 
\end{algorithm}

The idea of the reduction is to transform a Kappa system $\mathcal{R}$  to a Kappa system $\mathcal{R}'$ with fewer rules and fewer agents, while still capturing the observables and the relevant dynamics.
Our algorithm statically analyzes the rule-set, in search for one of the following mechanistic schemes:
\begin{itemize}
\item the \emph{modifier elimination} and \emph{similar rule composition}, that are the patterns amenable to the exact reduction (providing the equivalent rule-based model with fewer rules), as well as
\item the \emph{\mmgen},  \emph{\mm} and \emph{\fastdim} reductions, three interaction patterns amenable to approximate reduction based on time-scale separation (\secref{approximation}).
\end{itemize}

Recall that the generic framework for time-scale separation in biochemical reaction networks is shown in \cite{kang2013separation}.
A special case of this framework is \thmref{thm:kang}, which confirms that using the classical enzymatic approximation in stochastic setting is adequate.
%
%
After detecting one of the five interaction patterns, our algorithms, similarly as in \cite{kuwahara2006automated}, 
perform additional checks, in order to avoid the situations where the equilibrium assumptions are violated due to interleavings with the rest of the reaction network.

The top-level algorithm is shown in \algref{algo:Red}. 
We next describe each of the five interaction patterns in more detail.
 
\subsection{Similar rule composition}
In \emph{similar rule composition} scheme, rules have the same reactants, modifiers and products, but different rates. 
Our algorithm combines them into a single rule, by summing their rate laws.
Notice that this reduction is exact, that is, applying the similar rule composition does not change the semantics of the rule-based system.

\subsection{Modifier elimination}
This reduction can be applied when a species only appears as a modifier throughout a rule-based system.
Such a species will never change its copy number throughout the dynamics, 
and therefore, its quantity will be constant.
The species being always a modifier does affect the dynamics of the system, and all the rule rates where the species was involved need to be adapted.
Concretely -- after the species is eliminated, each rate law will be multiplied by the initial copy number of this species.
Notice that modifier elimination reduction is exact, that is, applying the modifier elimination does not change the semantics of the rule-based system.

\subsection{\Fastdim\;reduction}
The algorithm searches for dimerisation rules.
Suppose that a pair of reversible reactions $M+M\leftrightarrow M_2$ is detected.
%
Before proceeding to the reduction, we check whether a dimer is produced elsewhere, or if the monomer is a modifier elsewhere.
These checks are necessary because they prevent from deviating from the assumed equilibrium. 
Finally, if all checks passed, the dimerization reaction can be eliminated.
A new species $M_T$ is introduced, and, wherever the monomer $M$ or dimer $M_2$ were involved, they are replaced by the species $M_T$, and the rate is adapted accordingly, by the expressions shown in \secref{sec:dimerization}. 
%
\subsection{Generalised enzymatic reduction}
The algorithm searches for the scheme described in \secref{sec:generalized}, by searching for candidate enzymes.
Each pattern is tested as to whether it is catalyzing some enzymatic reduction. 
If a pattern $\specP$ indeed is an enzyme (operator) in an enzymatic reaction scheme, a set of all patterns $c$ which compete to bind to $\specP$ is formed, as well as the set of their complexes $sc$.
Then, before proceeding with the reduction, additional tests must be performed:
(i) pattern $\specP$ must be a species, and it is not an observable, 
(ii) $\specP$ must be small in copy number, that is, its initial copy number is smaller than a threshold,
(iii) $\specP$ can neither be produced, nor degraded,
(iv) complex $sc$ is not an observable and is never appearing in another rule of $\mathcal{R}$ and has initially zero abundance.
Then, the patterns $\specP$ and $\specP c$ can be eliminated from the rule-set and the reaction rates are adjusted according to the description in \secref{sec:generalized}.

%
%
Often times, \mm\;reduction is appropriate to eliminate the binding of the transcription factor to the operator site. 
In this context, the operator site takes the role of the enzyme, and transcription factor(s) the role of the substrate. 
Whenever a candidate enzyme is detected, and the other algorithm checks pass, the rates are appropriately scaled.
%
%
The competitive enzymatic reduction is suitable in a situation when more transcription factors compete for binding the enzyme, each in a different reaction.
In other words, the algorithm finds $k$ rules where $k$ different substrates compete for the same enzyme.

%

\begin{example} \rm

We illustrate the competitive enzymatic transformation on a small subnetwork of the $\lambda$-phage model, which will be introduced in \secref{results}.
The four rules presented below model the binding of the agent $\agent{RNAP}{}$ to the operator site of the agent $\agent{PRE}{}$ and subsequent production of protein $\agent{CI}{}$.
Agent $\agent{PRE}{}$ binds either only $\agent{RNAP}{}$ (at rate $\rate{1+}$ and $\rate{1-}$), or simultaneously with $\agent{CII}{}$ (at rate $\rate{2+}$ and $\rate{2-}$).
The protein can be produced whenever $\agent{PRE}{}$ and $\agent{PRE}{}$ are bound, but the rates will be different depending on whether only $\agent{RNAP}{}$ is bound to the operator (rate $\rate{b}$), or, in addition, $\agent{CII}{}$ is  bound to the operator (rate $\rate{a}$):

\begin{align*}
& \agent{PRE}{cii,rnap}, \agent{RNAP}{p1,p2} \\
&\leftrightarrow \agent{PRE}{cii,rnap!1},\agent{RNAP}{p1!1,p2} @  \rate{1+}, \rate{1-} \\
&\\
&
  \agent{PRE}{cii,rnap}, \agent{CII}{pre}, \agent{RNAP}{p1,p2}\\ 
  & \leftrightarrow \agent{PRE}{cii!1,rnap!2}, \agent{CII}{pre!1},  \agent{RNAP}{p1!2,p2} @ \rate{a+},\rate{a-} \\
  &\\
  &
 \agent{PRE}{cii,rnap!1}, \agent{RNAP}{p1!1,p2} \\& \rightarrow  \agent{PRE}{cii,rnap!1}, \agent{RNAP}{p1!1,p2}, 10\agent{CI}{ci,or} @ \rate{b}\\
 &\\
 &
  \agent{PRE}{cii!1,rnap!2}, \agent{CII}{pre!1}, \agent{RNAP}{p1!2,p2} \\&\rightarrow  \agent{PRE}{cii!1,rnap!2}, \agent{CII}{pre!1},\agent{RNAP}{p1!2,p2}, 10\agent{CI}{ci,or}@ \rate{a}
  \end{align*}
  
  After the competitive enzymatic reduction, the operator $\agent{PRE}{}$ is eliminated from each of the two competing \mm\; patterns. 
  Finally, the production of $\agent{CI}{}$ is modelled only as a function of $\agent{RNAP}{}$ and $\agent{CII}{}$, and the rate is appropriately modified:
  \begin{align*}
&\agent{RNAP}{p1,p2}, \agent{CII}{pre}\\& \rightarrow 
\agent{RNAP}{p1,p2}, \agent{CII}{pre}, 10\agent{CI}{pr,ci}
@ \;\; \rate{new}.
\end{align*}
\end{example}

\section{Case Study: $\lambda$-phage}
\label{results}
\begin{figure}[t]
\begin{center}
\includegraphics[scale=0.5]{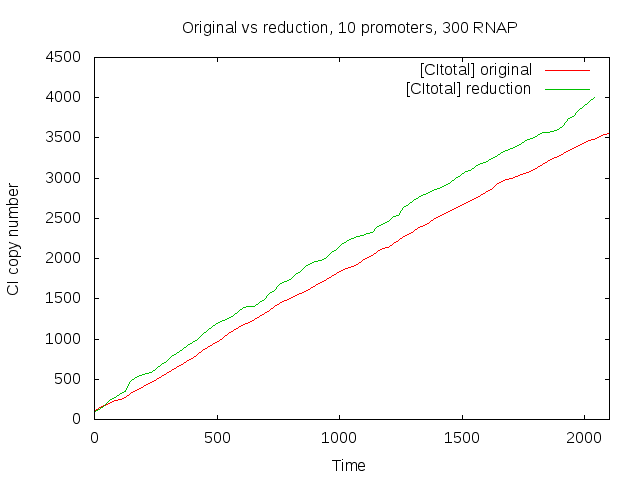}
\end{center}
\caption{Average trace of 10 simulations of the original model (red) and the reduced model (green) after the reduction, for initially $10$ $\lambda$ phage cells (multiplicities of infection -- MOI's). The simulation time for one simulation trace of the original model is $\approx 40$ minutes of CPU time, and of the reduced model is $5$ seconds of CPU time.
The initial number of proteins $\mathtt{CI}$, $\mathtt{Cro}$, $\mathtt{CII}$ and $\mathtt{CIII}$ and $\mathtt{N}$ is set to $100$.
}
\label{fig:simulations}
\end{figure}

\begin{figure}[t]
\begin{center}
\includegraphics[scale=0.5]{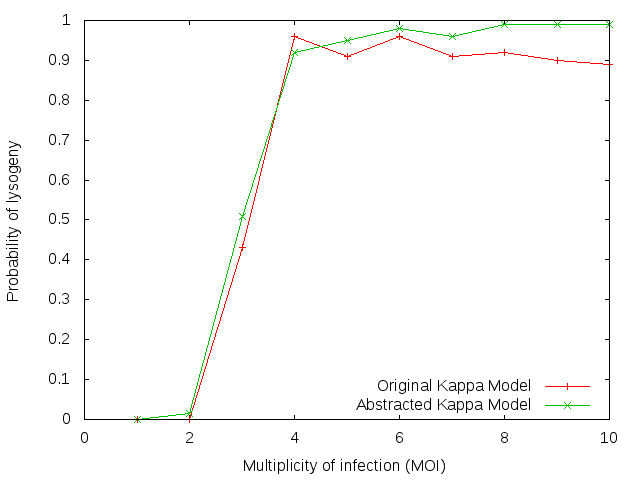}
\end{center}
\caption{Comparison of the probability of lysogeny before and after the reduction of the model (lysogeny profile is detected if there are $328$ molecules of \texttt{CI}  before there are $133$ molecules of \texttt{Cro}). 
The profile was obtained by running 1000 simulations of the model for one cell cycle (2100 time units), for MOIs ranging from 1 to 10. Simulation times are as reported in \figref{fig:simulations}.
}
\label{fig:lysogenyprob}
\end{figure}

The phage $\lambda$ is a virus that infects \emph{E.coli} cells, and replicates using one of the two strategies: \emph{lysis} or \emph{lysogeny}.
In the \emph{lysis} strategy, phage $\lambda$ uses the machinery of the \emph{E.coli} cell to replicate itself
and then lyses the cell wall, killing the cell and allowing the newly formed viruses to escape and infect other cells, while in the \emph{lysogeny} scenario, it inserts its DNA into the host cell's DNA and replicates through normal cell division, remaining in a latent state in the host cell (it can always revert to the lysis strategy).
The decision between \emph{lysis} and \emph{lysogeny} is known to be influenced by environmental parameters, as well as the multiplicity of infection and variations in the average phage input \cite{arkin1998stochastic}.

\begin{figure}[t]
\begin{center}
\includegraphics[scale=0.3]{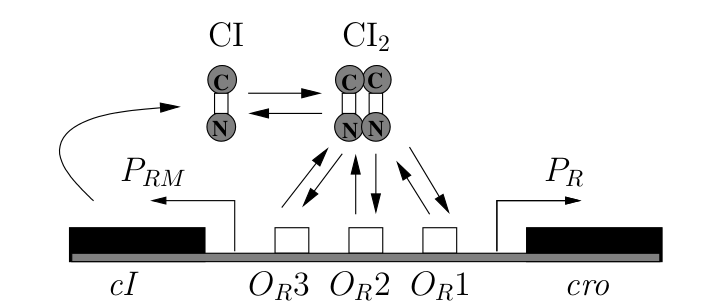}
\end{center}
\caption{CI monomers are produced from the cI gene; two monomers can form a dimer, that can bind to one of the $O_R$ operator sites (the Figure is taken from \cite{myers2011engineering}).}
\label{fig:CI}
\end{figure}

The key element controlling the decision process is the \emph{$O_R$} operator (shown in \figref{fig:CI}), which is composed of three operator sites ($O_{R1}$, $O_{R2}$, $O_{R3}$) to which transcription factors can bind, in order to activate or repress the two promoters ($P_{RM}$ and $P_{R}$) overlapping the operator sites. When RNAP (RNA polymerase, an enzyme that produces primary transcript RNA) binds to $P_{RM}$, it initiates transcription to the left, to produce mRNA transcripts from the \emph{cI} gene; RNAP bound to the $P_R$ promoter, on the other hand, initiates transcription to the right, producing transcripts from the \emph{cro} gene.  The two promoters form a genetic switch, since transcripts can typically only be produced in one direction at a time.

The \emph{cI} gene codes for the CI protein, also known as the \emph{$\lambda$ repressor}: in its dimer form (two CI monomers react to form a dimer, $CI_2$), it is attracted the the $O_R$ operator sites in the phage's DNA, repressing the $P_R$ promoter from which Cro production is initiated and further activating CI production.
Similarly, the \emph{cro} gene codes for the Cro protein, which also dimerizes in order to bind to $O_R$ operator sites and prevent production from $P_{RM}$, or even its own production.

While $CI_2$ and $Cro_2$ can bind to any of the three operator sites at any time, they have a different affinity to each site. The $CI_2$ has its strongest
affinity to the $O_{R1}$ operator site, next to the $O_{R2}$ site, and finally to the $O_{R3}$ site (in other words, $CI_2$ first turns off $P_R$ , then activates $P_ {RM}$ , and finally, represses its own production), while $Cro_2$ has the reverse affinity (it first turns off CI production, then turns off its own production).

The feedback through the binding of the products as transcription factors coupled with the affinities described makes the $O_R$ operator behave as a genetic bistable switch. In one state, Cro is produced locking out production of CI. In this state, the cell follows the \emph{lysis} pathway since genes downstream of Cro produce the proteins necessary to construct new viruses and lyse the cell. In the other state, CI is produced locking out production of Cro. In this state, the cell follows the \emph{lysogeny} pathway since proteins necessary to produce new viruses are not produced. Instead, proteins to insert the DNA of the phage into the host cell are produced.

What's more, in the lysogeny state, the cell develops an immunity to further infection: the cro genes found on the DNA inserted by further infections of the virus
are also shut off by $CI_2$ molecules that are produced by the first virus to commit to lysogeny. Once a cell commits to lysogeny, it becomes very stable and does not easily change over to the lysis pathway. An induction event is necessary to cause the transition from lysogeny to lysis. For example, lysogens (i.e., cells with phage DNA integrated within their own DNA) that are exposed to UV light end up following the lysis pathway.

\subsection{Results and discussion}

We applied our reduction algorithm to a Kappa model of the phage $\lambda$ decision circuit that we built using the reaction-based model presented in \cite{myers2011engineering}, \cite{kuwahara2006automated}.

\subsubsection{Intermediate tests}


Intermediate tests were carried out on the portion of the circuit that is involved in CI production from the he $P_{RE}$ promoter and in CII production from the $P_R$ promoter. 
Initially, this model contained 10 rules, 5 proteins and 4 species; after applying the reduction algorithm, it was reduced to 4 rules and 3 proteins.


%

\subsubsection{Full model}

\begin{figure}[t]
\begin{center}
\includegraphics[scale=0.3]{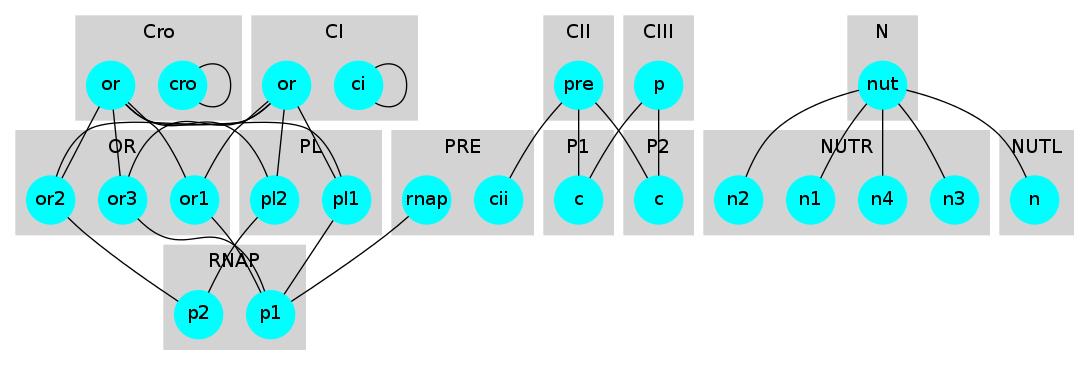}
\end{center}
\caption{The contact map of the full $\lambda$-phage model. The model consists of  $96$ rules, $16$ proteins and $61$ species.
The reduced model has $11$ rules and $5$ proteins. 
}
\label{fig:CM}
\end{figure}

Simulations were carried out on 
the complete chemical reaction genetic circuit model which contains 96 rules, 16 proteins and 61 species (the contact map is shown in \figref{fig:CM}). After applying the reduction, the Kappa model is reduced to 11 rules and 5 proteins. 

%
In \figref{fig:simulations}, we plot the mean for the $\texttt{CI}$ copy number obtained from $10$ runs of the original and of the reduced model, and the graphs show agreement.
%
%

%
In \figref{fig:lysogenyprob}, we compared the probability of lysogeny before and after the reduction of the model (lysogeny profile is detected if there are $328$ molecules of \texttt{CI}  before there are $133$ molecules of \texttt{Cro}).
The graphs show overall agreement in predicting the lysogeny profile.
More precisely, for two and less MOI's (multiplicities of infection), the probability of lysogeny is almost negligible; For three MOIs, both graphs show that lysogeny and lysis are equally probable (the reduced model reports slightly larger probability),
and for five or more MOI's, both graphs show that lysogeny is highly probable.
While one simulation of the original model takes about 40 mins, one simulation of the abstracted model takes about 5 seconds. Once again, the results are similar, with a significant improvement in simulation speed. 

%

The tool is available for download \cite{webKaRed}.

%
%




\section{Conclusion and future work}
In this paper, we presented a method for automated reduction of rule-based models. 
The reduction is based on equilibrium approximations: certain rules and species are eliminated and the rates are approximated accordingly. 
More concretely, a number of reaction patterns known to be amenable to equilibrium approximations are recognised by static inspection of the rules.
%
%
The crucial aspect of the presented approach is that each approximation step can be retrieved at any time, and no information about the original, detailed model is lost. 
The presented method can be seen as the first step towards a systematic time-scale separation of stochastic rule-based models.
The guarantees of the presented reduction method are given for the asymptotic behaviour.
Bhattacharyya distance is proposed as a metric to quantify the reduction error with respect to the observable.
We plan to further investigate how to practically access the approximation error.
To this end, the error can be measured with respect to a given observable, or, more generally, with respect to a given property specified in, for example, linear temporal logic (LTL).
%

\begin{figure}[t]
\begin{center}
\includegraphics[scale=0.25]{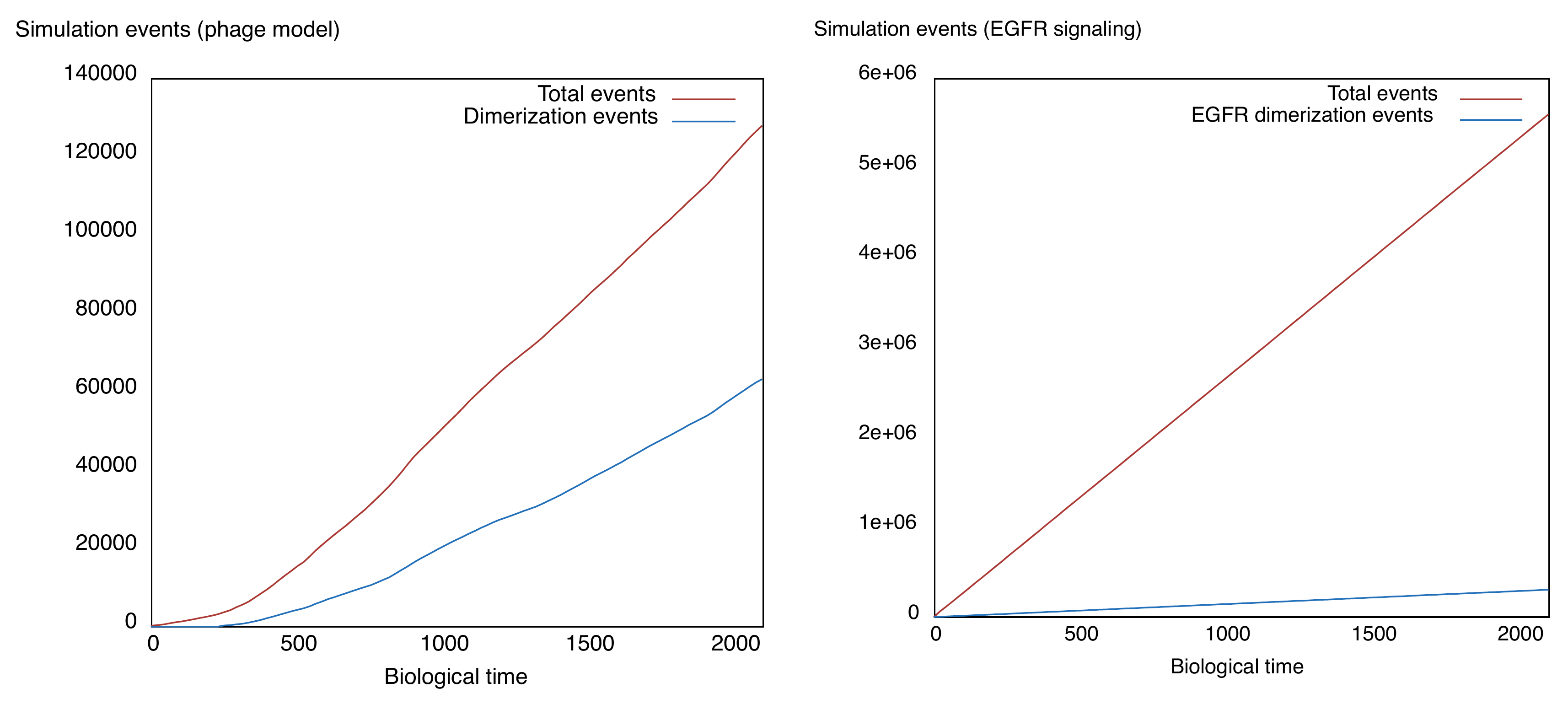}
\end{center}
\caption{
a) 
The ratio of dimerisation events vs. total events in lambda phage model.
The number of dimerisation events takes roughly half of the total events over the whole cell cycle.
b) 
The ratio of dimerisation events vs. total events in EGFR/insulin model.
The number of dimerisation events takes only a small fraction of the total events over the whole cell cycle.
}
\label{fig:eventratio}
\end{figure}

%
We implemented the tool and evaluated it on a case study of a lambda phage bistable switch. 
%
The simulation of one cell cycle was improved from $40min$ CPU time to $5sec$, and the profiles of the observables show agreement.
We plan to extend the set of approximation patterns so to obtain good reductions for complex models of signaling pathways. 
More precisely, while our tool is applicable to any rule-based model, the chosen set of approximation patterns are tailored for GRNs and may not provide significant reductions when applied to the signaling pathways.
To illustrate this, we applied the reduction to the EGF/insuling crosstalk model, and we observe that the number of dimerisation events does not take the significant portion of all events (see \figref{fig:eventratio}),
%
at least not as radically as it was the case with the lambda phage example. 
To this end, we plan to include more patterns for reducing signaling pathways, by, for example, approximating multiple phosphorylation events.

 \section*{Acknowledgements} 
 The authors would like to thank to J\'er\^{o}me Feret, for the inspiring discussions and useful suggestions.

\bibliographystyle{abbrv}
\bibliography{MyRef.bib}  
%
%

%

\balancecolumns
\end{document}